\documentclass[prb,showpacs,twocolumn,aps,superscriptaddress,a4paper]{revtex4}
\usepackage{dcolumn}
\usepackage{amsmath}
\usepackage{graphicx}
\usepackage{latexsym}  
\usepackage{amsfonts}  
\usepackage{amssymb}  
\DeclareGraphicsExtensions{.eps,.pdf,.gif,.jpg}  
  
\newcommand{\be}{\begin{equation}}  
\newcommand{\ee}{\end{equation}}
\newcommand{\beq}{\begin{eqnarray}}  
\newcommand{\eeq}{\end{eqnarray}}

\begin{document}  
      
\def\bbe{\mbox{\boldmath $e$}}  
\def\bbf{\mbox{\boldmath $f$}}      
\def\bg{\mbox{\boldmath $g$}}  
\def\bh{\mbox{\boldmath $h$}}  
\def\bj{\mbox{\boldmath $j$}}  
\def\bq{\mbox{\boldmath $q$}}  
\def\bp{\mbox{\boldmath $p$}}  
\def\br{\mbox{\boldmath $r$}}      
  
\def\bone{\mbox{\boldmath $1$}}      
  
\def\dr{{\rm d}}  
  
\def\tb{\bar{t}}  
\def\zb{\bar{z}}  
  
\def\tgb{\bar{\tau}}

\def\bC{\mbox{\boldmath $C$}}  
\def\bG{\mbox{\boldmath $G$}}  
\def\bH{\mbox{\boldmath $H$}}  
\def\bK{\mbox{\boldmath $K$}}  
\def\bM{\mbox{\boldmath $M$}}  
\def\bN{\mbox{\boldmath $N$}}  
\def\bO{\mbox{\boldmath $O$}}  
\def\bQ{\mbox{\boldmath $Q$}}  
\def\bR{\mbox{\boldmath $R$}}  
\def\bS{\mbox{\boldmath $S$}}  
\def\bT{\mbox{\boldmath $T$}}  
\def\bU{\mbox{\boldmath $U$}}  
\def\bV{\mbox{\boldmath $V$}}  
\def\bZ{\mbox{\boldmath $Z$}}  
  
\def\bcalS{\mbox{\boldmath $\mathcal{S}$}}  
\def\bcalG{\mbox{\boldmath $\mathcal{G}$}}  
\def\bcalE{\mbox{\boldmath $\mathcal{E}$}}  
  
\def\bgG{\mbox{\boldmath $\Gamma$}}  
\def\bgL{\mbox{\boldmath $\Lambda$}}  
\def\bgS{\mbox{\boldmath $\Sigma$}}  
  
\def\bgr{\mbox{\boldmath $\rho$}}  
  
\def\a{\alpha}  
\def\b{\beta}  
\def\g{\gamma}  
\def\G{\Gamma}  
\def\d{\delta}  
\def\D{\Delta}  
\def\e{\epsilon}  
\def\ve{\varepsilon}  
\def\z{\zeta}  
\def\h{\eta}  
\def\th{\theta}  
\def\k{\kappa}  
\def\l{\lambda}  
\def\L{\Lambda}  
\def\m{\mu}  
\def\n{\nu}  
\def\x{\xi}  
\def\X{\Xi}  
\def\p{\pi}  
\def\P{\Pi}  
\def\r{\rho}  
\def\s{\sigma}  
\def\S{\Sigma}  
\def\t{\tau}  
\def\f{\phi}  
\def\vf{\varphi}  
\def\F{\Phi}  
\def\c{\chi}  
\def\w{\omega}  
\def\W{\Omega}  
\def\Q{\Psi}  
\def\q{\psi}  
  
\def\ua{\uparrow}  
\def\da{\downarrow}  
\def\de{\partial}  
\def\inf{\infty}  
\def\ra{\rightarrow}  
\def\bra{\langle}  
\def\ket{\rangle}  
\def\grad{\mbox{\boldmath $\nabla$}}  
\def\Tr{{\rm Tr}}  
\def\Re{{\rm Re}}  
\def\Im{{\rm Im}}

\title{Bound States in Time-Dependent Quantum Transport: Oscillations and 
Memory Effects in Current and Density} 
 
\author{E. Khosravi}  
\affiliation{Institut f\"ur Theoretische Physik, Freie Universit\"at Berlin,   
Arnimallee 14, D-14195 Berlin, Germany}  
\affiliation{European Theoretical Spectroscopy Facility (ETSF)} 
\author{G. Stefanucci}   
\affiliation{Department of Physics, University of Rome Tor Vergata, 
Via della Ricerca Scientifica 1, 00133 Rome, Italy}
\affiliation{European Theoretical Spectroscopy Facility (ETSF)}  
\author{S. Kurth}  
\affiliation{Institut f\"ur Theoretische Physik, Freie Universit\"at Berlin,   
Arnimallee 14, D-14195 Berlin, Germany}  
\affiliation{European Theoretical Spectroscopy Facility (ETSF)}    
\author{E.K.U. Gross}  
\affiliation{Institut f\"ur Theoretische Physik, Freie Universit\"at Berlin,   
Arnimallee 14, D-14195 Berlin, Germany}  
\affiliation{European Theoretical Spectroscopy Facility (ETSF)}

\date{\today}  

\begin{abstract}
The presence of bound states in a nanoscale electronic system attached to 
two biased, macroscopic electrodes is shown to give rise to persistent, 
non-decaying, localized current oscillations  
which can be much larger than the steady part of the current. The amplitude 
of these oscillations depends on the entire history 
of the applied potential. The bound-state contribution to the {\em static} 
density is history-dependent as well. Moreover, the time-dependent formulation 
leads to a natural definition of the bound-state occupations out of 
equilibrium.
\end{abstract}
  
  
\maketitle  
  
In recent years it has become possible to measure the current through single 
molecules attached to two macroscopic electrodes 
\cite{ReedZhouMullerBurginTour:97,YansonBollingerBromAgraitRuitenbeek:98}. 
This is hoped to be a first step towards the vision of ``Molecular 
Electronics'' where single molecules become the basic units (transistors, 
etc.) of highly miniaturized electronic devices. 

To address electronic transport on such a small scale theoretically, 
one needs a full quantum description of the electronic dynamics. 
Non-equilibrium Green's functions (NEGF) provide a 
natural framework to study quantum transport properties of nanoscale devices 
coupled to leads.  When a bias is applied, the electrodes   
remain in local equilibrium while the current is driven by the 
different chemical potentials in the left and right lead.
In model systems the leads are assumed to be 
non-interacting and the current is computed from the Meir-Wingreen
formula\cite{MeirWingreen:92} using approximate many-body self-energies $\S_{\rm MB}$. For weakly 
correlated models $\S_{\rm MB}\sim 0$ and the Meir-Wingreen
formula reduces to the Landauer-B\"uttiker formula\cite{Landauer:57,Buettiker:86}, as it should.

If one wants to account for the full atomistic structure of the 
system,
the NEGF formalism is usually combined with static density functional theory (DFT) and 
the current is computed from a Landauer-type equation
\cite{Lang:95,ht.1995,SeminarioZacariasTour:98,tgw.2001,pplv.2001,XueDattaRatner:02,BrandbygeMozosOrdejonTaylorStokbro:02,EversWeigendKoentopp:04,flss.2005}. 
This approach enjoys increasing popularity, in particular for the description 
of transport experiments on single molecules \cite{ReedZhouMullerBurginTour:97}. 
From a fundamental point of view, however, the use of static DFT  - which is 
an equilibrium theory - is not justified to describe  non-equilibrium 
situations. For a critical review of this methodology, the reader is referred 
to Ref.~\onlinecite{KoentoppChangBurkeCar:08}.

By construction, the NEGF+DFT approach inherits the 
main assumption of the Landauer formalism that  
for a system driven out of equilibrium by a dc 
bias, a steady current will eventually develop. In other words,
the dynamical formation of a steady state does not follow from the 
formalism but rather constitutes an assumption.
The question of how the system reaches the steady state has been 
addressed theoretically in 
Refs.~\onlinecite{StefanucciAlmbladh:04,StefanucciAlmbladh:04-2} where it was 
shown that the total current (and the density) reaches a steady value if the 
local density of states is a smooth function of energy in the device region. 
In the same work it was also shown that for non-interacting electrons 
the steady current is 
independent both of the initial conditions and of the history of the bias, 
i.e., all memory effects are washed out. 

The situation is different, however, if there exist two or more localized 
bound states (BS) in the device region, i.e., if the local density of states 
has sharp peaks at certain energies. This case has been investigated in 
Ref.~\onlinecite{DharSen:06} and further been elaborated in 
Ref.~\onlinecite{Stefanucci:07} where it was shown analytically 
that a non-interacting system with bound states exposed to a dc 
bias {\em does not} evolve to a steady state. Instead, 
a steady component of the current is superimposed by undamped harmonic 
current oscillations: Assuming that the one-particle Hamiltonian $\bH(t)$ 
globally converges to a time-independent Hamiltonian $\bH^{\inf}$ when 
$t\ra\inf$, the total current through a plane $\P$ perpendicular to the 
transport geometry has the form 
\begin{equation}
\lim_{t\rightarrow \inf} 
I_{\P}(t)=I_{\P}^{(S)}+I_{\P}^{(D)}(t) \; ,
\label{s+dcurr} 
\end{equation} 
where the static contribution $I_{\P}^{(S)}$ is given by the Landauer formula.  
The dynamical part, $I_{\P}^{(D)}(t)$, reads (atomic units are used throughout)
\begin{equation}
I_{\P} ^{(D)}(t)= 2\sum_{b,b'} {f_{b,b'} \L _{b,b'} ^{\P} 
\sin[(\epsilon_b^{\inf} -\epsilon_{b'}^{\inf})t]},
\label{dcurr} 
\end{equation} 
where the summation is over all BS of the final Hamiltonian 
$\bH^{\inf}$ and the oscillation frequencies are
given by the BS eigenenergy differences. The 
quantities ${\L}^{\P}_{b,b'}$ and $f_{b,b'}$ are defined according to  
\begin{equation}
\L_ {b,b'} ^{\P}=
\int_{\P}\dr\s\dr\s'
\q^{\inf}_{b'}(\br')\S(\br',\br;\e^{\inf}_{b'})\q^{\inf}_{b}(\br),
\label{lamda}
\end{equation} 
and 
\begin{equation}
f_{b,b'}= \bra \psi'_{b }|f(\bH^{0})|\psi'_{b' } \ket \; .
\label {fbb'}
\end{equation}
In Eq. (\ref{lamda}) the double surface integral is over the plane $\P$, 
$\S$ is the embedding self-energy and $\q^{\inf}_{b}(\br)$ 
are BS eigenfunctions. The operator $f(\bH^{0})$ in Eq. (\ref{fbb'})
is the Fermi function calculated at the equilibrium Hamiltonian $\bH^{0}$
while the state  
$|\psi'_{b } \ket$ is related to $|\psi_{b }^{\inf} \ket$ by a 
unitary transformation: $|\psi'_{b } \ket=\bM |\psi_{b }^{\inf} \ket$.
The ``memory operator'' $\bM$ depends on the history of the 
TD perturbation. 
Therefore the coefficients $f_{b,b'}$ are  
matrix elements of $f(\bH^{0})$ between history-dependent 
localized functions.

In the present work we study these time-dependent (TD) phenomena for a 
number of numerical examples using a recently proposed algorithm 
for TD  transport \cite{KurthStefanucciAlmbladhRubioGross:05}. We 
show numerically that the amplitude of the current oscillations 
may be very large compared to the steady component of the current. 
In addition, the memory dependence of the current oscillations will be 
explicitly demonstrated. Furthermore, we address an important aspect which, 
in the NEGF+DFT approach, has remained an open problem, namely how to take 
BS into account in the calculation of the density
\cite{XueDattaRatner:02,KeBarangerYang:04}. Recently, a somewhat empirical 
scheme was suggested on how to occupy BS with energies in the bias 
window \cite{LiZhangHouQianShenZhaoXue:07}. Here we will show how the BS 
occupations naturally {\em result from the time evolution} of the 
underlying KS system. Moreover, for non-interacting electrons 
we provide numerical evidence 
that these occupations (and therefore the density in the device region) show 
a history dependence as well.

In the long-time limit, the dynamical contribution to the density is given by 
\begin{equation}
n^{(D)}(\br, t) = \sum_{b,b'} {f_{b,b'} 
\cos[(\epsilon_b^{\inf} -\epsilon_{b'}^{\inf})t] \psi_{b}^{\inf}(\br)
\psi_{b'}^{\inf}(\br)},
\label{dyndens}
\end{equation} 
where the amplitudes $f_{b,b'}$ are again given by Eq.~(\ref{fbb'}). 
We observe that while the diagonal term, $b=b'$, 
does not contribute to the current of Eq.~(\ref{dcurr}), it does contribute 
to the density of Eq.~(\ref{dyndens}) with a history-dependent coefficient 
$f_{b,b}$. This means that even if we average out the density oscillations,
history dependent effects will show up in the density at the device region. 
This is a central result of our analysis. 

In order to shed more light on the actual dependence of the 
coefficients $f_{b,b'}$ on the history of the TD 
perturbation  we study model systems of non-interacting electrons 
using a recently proposed algorithm 
\cite{KurthStefanucciAlmbladhRubioGross:05} suitable to describe 
TD transport through open systems. The 
central feature of this algorithm is that it allows for the numerically exact 
solution of the TD Schr\"odinger equation in the central device 
region in the presence of the semi-infinite electrodes. 

We consider one-dimensional systems described by the TD 
Hamiltonian
\begin{equation}  
    H(x,t)=-\frac{1}{2} \frac{{\rm d}^{2}}{{\rm d}x^2}+U_0(x)+U(x,t) \;.  
\end{equation}
For times $t<0$ the system is in its ground state described by the Hamiltonian 
$H^{0}(x)=-\frac{\nabla^{2}}{2}+U_{0}(x)$. At $t=0$ the system is driven out 
of equilibrium by applying a TD field $U(x,t)$. We choose the 
TD perturbation in such a way that for ${t\rightarrow \inf}$ the 
Hamiltonian globally converges to an asymptotic Hamiltonian, $H^{\inf}$. 

The TD perturbation can be split into three sub-regions depending 
on where in the system it is imposed: $U_{\a}$, $\a=L,R$, represents the 
applied bias in the left ($L$) and right ($R$) leads and is independent 
of the position within the 
lead. In the central region, the TD perturbation (which we call 
a gate voltage $V_g(x,t)$) may depend both on position $x$ and time $t$. Thus, 
the total TD perturbation can be written as
 \begin{equation}
       U(x,t) = \left \{ \begin{array} {cc}
       U_{L}(t) &       -\inf < x < x_{L}\\
       V_g(x,t)  &       x_L < x < x_{R}\\
       U_{R}(t)   &    x_{R}< x < \inf
         \end{array}\right..
 \end{equation}
In the numerical simulations described below, the explicit propagation 
window ranges from $x_L=-1.2$ a.u. to $x_R=1.2$ a.u.. We choose a lattice 
spacing $\D x=0.012$ a.u. and 
use a simple three-point discretization for the kinetic energy. 
The initial potential is $U_0(x)=0$ for any point $x$ in the left 
or right lead. Therefore, the occupied part of the continuous spectrum ranges 
from momenta $k=-k_{\rm F}=-\sqrt{2\ve_{\rm F}}$ to $k=k_{\rm F}$ which is 
discretized with 400 $k$-points ($\ve_{\rm F}$ being the Fermi energy). 
The (non-interacting) many-body state is 
propagated from $t=0$ to $t=1400$ a.u. using a time step $\D t=0.05$ a.u..

In the first model the initial 
state is a Slater determinant of plane waves with energies 
less than $\ve_{\rm F}=0.1$ a.u.. At $t=0$ we suddenly 
switch on a bias $U_L = 0.15$ a.u. in the left lead and as a result a current 
flows which after some transient time reaches a steady value of about $0.027$ 
a.u.. After the steady state is attained, at time $T=100$ a.u., we 
switch on a gate potential 
\begin{equation}
V_g(x,t)= -\frac{(t-T)}{T_g} v_g
\label{gatepot}
\end{equation} 
for $T<t<T+T_g$ with a switching time $T_g=20$ a.u. and the final depth of the 
potential well $v_g=1.3$ a.u.. For times $t\geq T+T_g$ this gate potential 
remains unchanged at $V_g(x,t)= -v_g$ and supports two BS at 
$\ve_1^{\inf} = -0.933$ a.u. and $\ve_2^{\inf}=-0.063$ a.u.. 
\begin{figure}[tb] 
\includegraphics[angle=0,width=0.4\textwidth,clip]{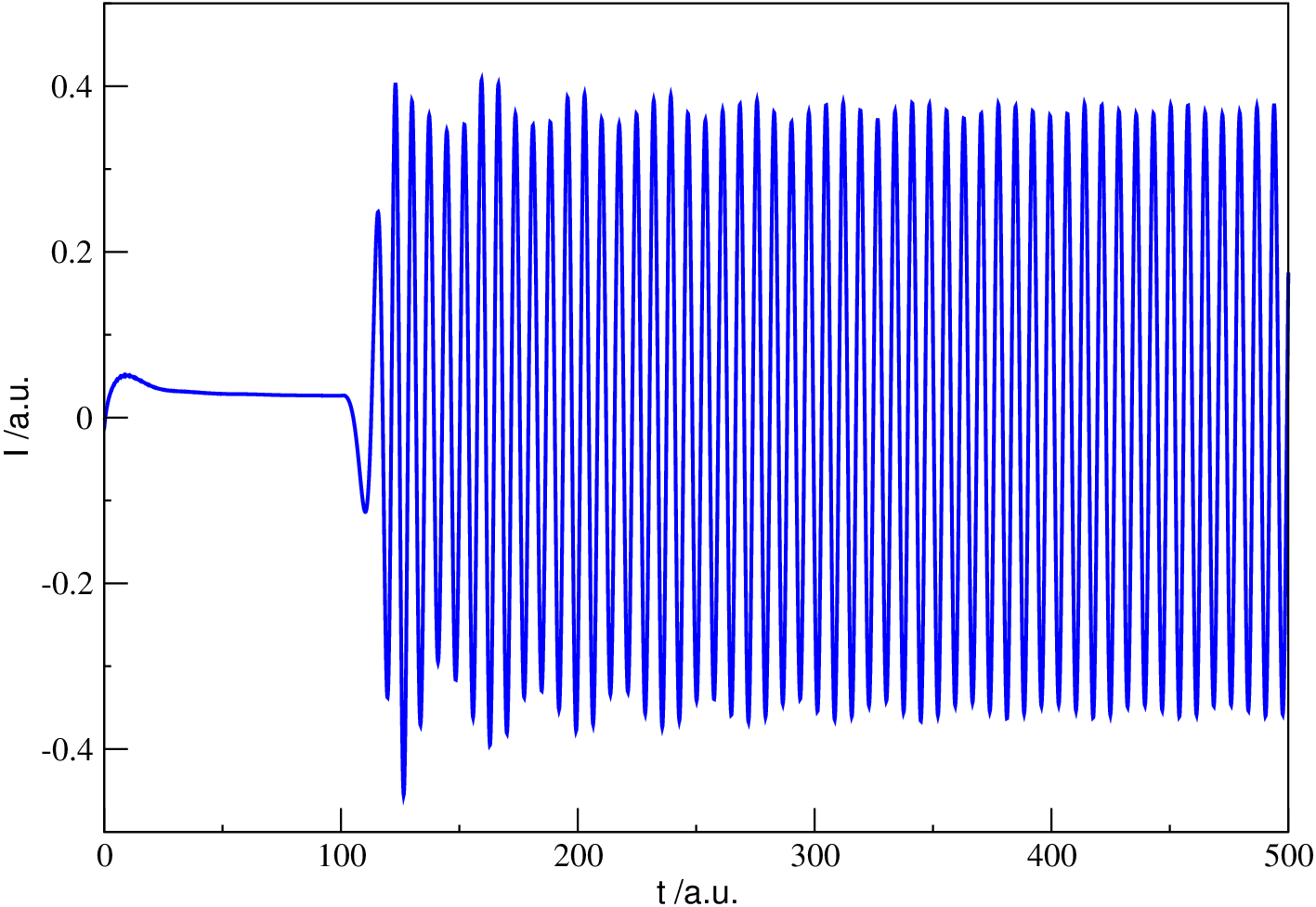}
\includegraphics[angle=0,width=0.4\textwidth,clip]{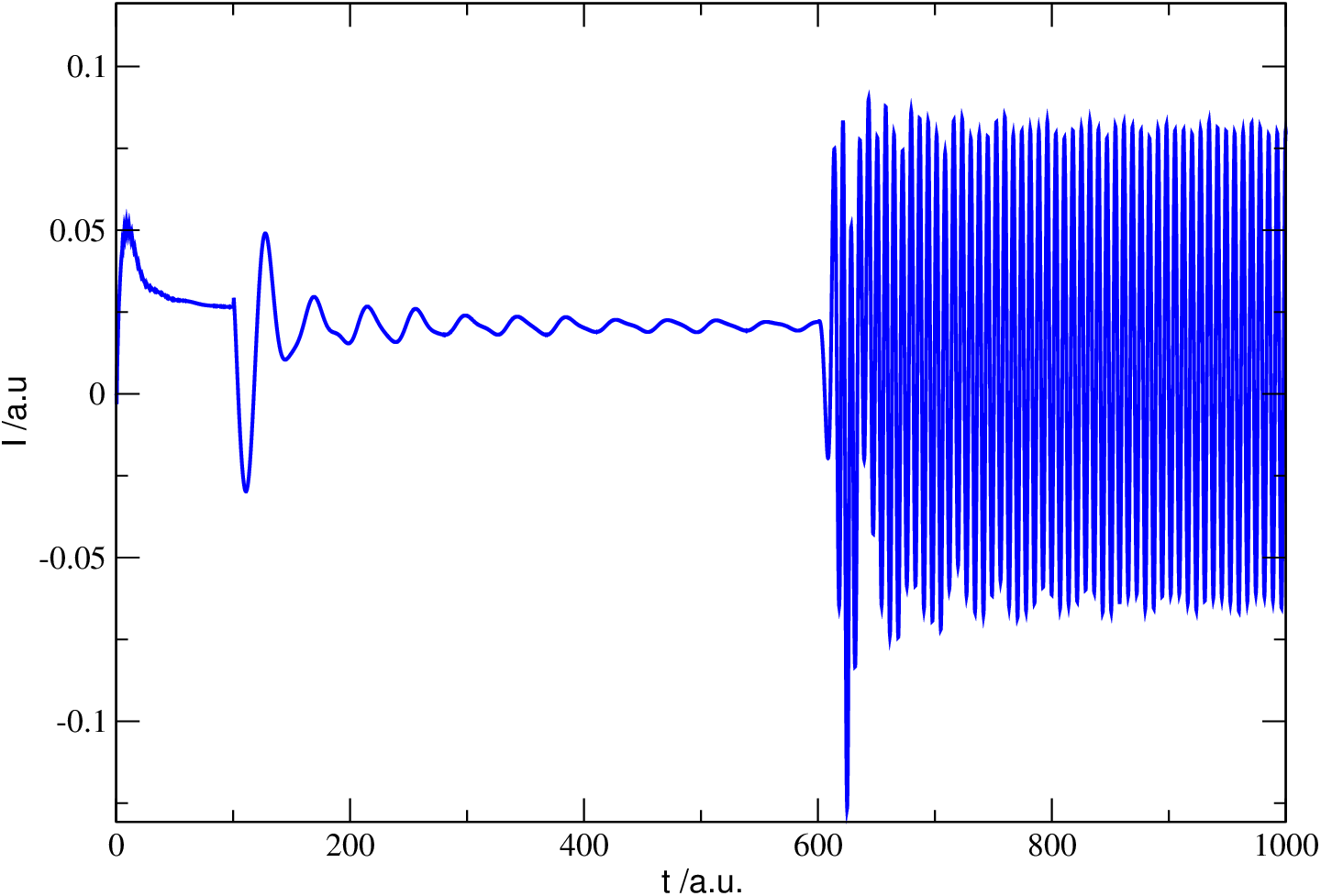}
\caption{Time evolution of the current at $x=0$. At $t=0$ a.u., 
a bias $U_L=0.15$ a.u. is suddenly switched on and the system 
evolves to a steady state. Upper panel: at $T=100$ a.u., a gate potential is 
turned on ($v_{g}=1.3$ a.u. and $T_{g}=20$ a.u.) which creates two BS 
and results in large amplitude oscillations of the current. Lower panel: 
at $T=100$ a.u., a first gate potential ($v_{g}=0.2$ a.u. and 
$T_{g}=0$) is turned on which creates 
a single BS. Waiting for the transients to 
decay, a second gate voltage ($v_{g}=1.1$ a.u. and $T_{g}=20$ a.u.) 
is then applied which leads to the formation of 
a second BS and therefore to persistent current oscillations. 
Although $\bH^{\inf}$
is identical in both cases, the amplitude of the current oscillation  
is significantly smaller in the second case, illustrating its dependence 
on the history of the system.}
\label{tdcurr}     
\end{figure}
The resulting TD current in the center of the device region is 
shown in the upper panel of Fig.~\ref{tdcurr}. The development of a 
steady-state current for $T<100$ a.u. can clearly be recognized. After the 
BS are created ($t>T+T_{g}$) the current starts 
to oscillate as expected. The amplitude of the current oscillation is of the 
order of 0.35 a.u., i.e., {\em more than an order magnitude larger than the 
steady-state current}. 
By Fourier transforming the TD current one can identify various 
transitions which contribute to the oscillating behavior. In the 
long-time limit, only the transition between BS survives. 
In the transient regime, however, one finds additional, 
power-law decaying ($1/t$) transitions from BS to the right continuum 
at $\ve_{\rm F}$ as well as to the left continuum at 
$\ve_{\rm F}+U_L$ 
\cite{StefanucciKurthRubioGross:08}. 

The history dependence of the current oscillations can be seen by comparing 
the current in the upper and lower panels of Fig.~\ref{tdcurr}. Both currents 
were computed by starting from the same initial state and applying the same 
bias at $t=0$. In the lower panel we create the same, final potential as in 
the upper panel, but in two steps. At $T=100$ a.u. we suddenly switch on a 
first gate potential with depth $v_{g}=0.2$ a.u. which creates one BS. Waiting 
for the slow decay of the resulting bound-continuum transition we then apply 
an additional gate potential of depth $v_{g}=1.1$ a.u. (hence the 
total depth is 1.3 a.u.) with a switching time 
$T_{g}= 20$ a.u.. Again we recognize the persistent current oscillations due 
to the bound-bound transitions. Although the amplitude 
(about 0.07 a.u.) is still large compared to the steady-state current, it is 
about a factor of four smaller than the amplitude in the previous 
case. 

Memory effects not only appear in the amplitude of the current oscillations  
but also in the BS contribution to the density $n^{(D)}$ (Eq.~(\ref{dyndens})) 
through the history dependence of the coefficients $f_{b,b'}$ of 
Eq.~(\ref{fbb'}). In the long-time limit, BS lead to 
oscillations in $n^{(D)}$ (which are connected to the 
current oscillations through the continuity equation) and also 
contribute to the steady part of $n^{(D)}$. This contribution is given by 
the diagonal part ($b=b'$) of the double sum in Eq.~(\ref{dyndens}) which 
implies that $f_{b,b}$ may be interpreted as 
occupation numbers of the BS in the long-time limit. In this way, 
the TD description provides a natural way to include BS
in a transport calculation. In the framework of the Landauer 
formalism this can only be achieved in a somewhat artificial way 
\cite{LiZhangHouQianShenZhaoXue:07}. 

\begin{figure}[tb] 
\includegraphics[angle=0,width=0.45\textwidth,clip]{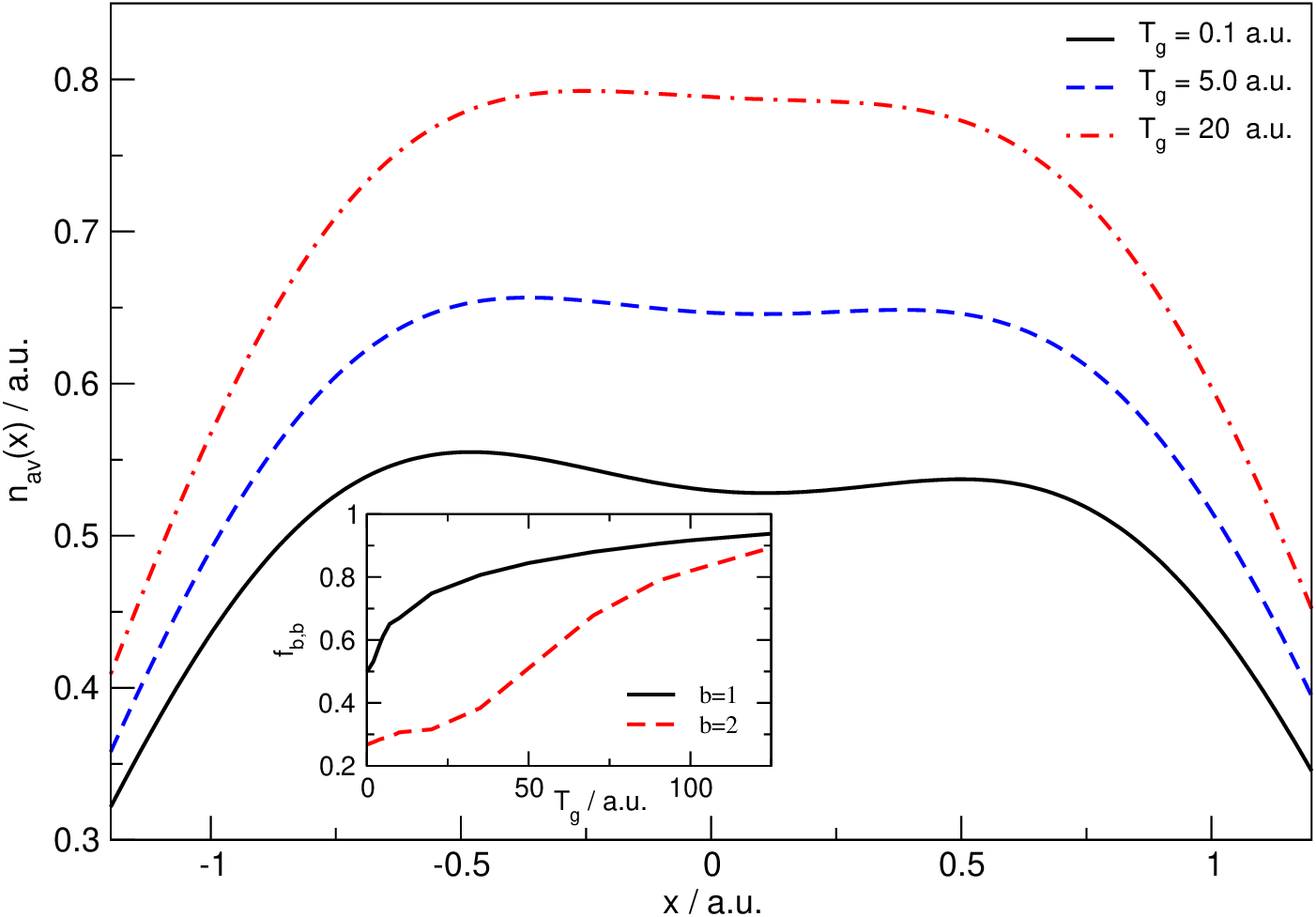}
\caption{Memory effects for the static part of the density in the long-time 
limit in the presence of BS. The densities shown here correspond to 
systems which are identical to the one studied in the upper panel of 
Fig.~\protect{\ref{tdcurr}} except that they are computed for three 
different switching times of the gate potential of 
Eq.~(\protect{\ref{gatepot}}): $T_g=0.1$ a.u. (solid, black), $T_g=5.0$ a.u. 
(dashed, blue), and $T_g=20$ a.u. (dash-dotted, red). The inset shows the 
occupation numbers $f_{b,b}$ (Eq.~(\protect\ref{fbb'})) 
of the two bound states as function of switching time. The state with lower 
energy eigenvalue ($b=1$) has higher occupation than the one with higher 
energy. For short switching times the occupation is significantly smaller 
than one, while for adiabatic (slow) switching both occupation numbers 
approach one. }
\label{memorydens}     
\end{figure}

We emphasize that also the diagonal occupations $f_{b,b}$ 
depend on the history through the memory operator $\bM$. 
Here we address the importance of this memory dependence which is 
rather crucial in the NEGF+DFT approach since BS below the bias window 
are entirely populated, i.e., $f_{b,b}=1$.
We have computed the time averaged density $n_{\rm av}(x)$ over an oscillation 
period for a system with two bound states in the final Hamiltonian.
The system is the one which leads to the 
current shown in the the upper panel of Fig.~\ref{tdcurr} except that the 
gate potential is turned on with different switching times $T_{g}$. 
In Fig.~\ref{memorydens} we show $n_{\rm av}(x)$  
for three different $T_{g}$ and, as one can see, $n_{\rm av}$
{\em differs quite substantially}. This difference has to be 
attributed to BS since for the contribution of the scattering states all 
memory is washed out \cite{StefanucciAlmbladh:04-2}. Of course, the relative 
importance of the BS contribution to $n_{\rm av}(x)$ will decrease if 
$\ve_{\rm F}$ (and therefore the contribution of the continuum states) 
increases. 
By subtracting the continuum contribution from $n_{\rm av}$
and fitting the explicit function $n^{(D)}$ with the numerical curve 
(where the only fitting parameters are the coefficients $f_{b,b}$ for $b=1,2$)
we have been able to calculate the BS occupations $f_{b,b}$. Remarkably 
we have found that $f_{b,b}$ {\em exhibits rather large deviations from 
unity}, particularly for small switching times (see inset of 
Fig.~\ref{memorydens}). As one would intuitively expect, the occupation of 
the state with lower energy ($b=1$) is larger than for the state with higher 
energy. In the adiabatic limit of very slow switching ($T_g \to 
\infty$), both occupation numbers approach unity which is expected 
since both states are energetically below $\ve_{\rm F}$. The occupation 
numbers also offer an intuitive qualitative picture of the size of the 
current and density oscillations: for relatively short switching times 
($T_g \leq 50$ a.u.) the occupation numbers deviate substantially from one. 
Therefore 
the transition probability between the bound states is relatively large 
and so are the oscillations in current and density. On the other hand, for 
large switching times both bound states are almost ``fully'' occupied and 
the probability of a transition between them is small, leading to small 
amplitudes in the dynamical density.

We have noticed that in the limit of slow switching the (transient) 
transitions between bound states and continuum can have a rather large 
amplitude shortly after the gate potential is applied, even larger than 
the (persistent) transition between the bound states. As a consequence of the 
very slow decay of the bound-continuum transitions, one has to propagate for a 
rather long time before the dynamical part of the density takes on the form of 
Eq.~(\ref{dyndens}). 

In summary we have demonstrated that 1) the persistent current oscillations in 
the presence of BS can be much larger than 
the steady-state current 2) the amplitude of the current oscillations can 
have a strong  dependence on the history of the system 3)
a similar history dependence is found both for the 
static and dynamic contribution of the BS to the density and 4) 
the occupation of the BS is well defined in a 
TD description of quantum transport.

In the calculations of the present work we assume the electrons to be 
non-interacting. However, the central conclusions (bound-state oscillations 
in current and density, memory effects in both quantities) remain valid for 
any effective single-particle theory such as, e.g., TD Hartree-Fock or TDDFT. 
In this case, the existence of BS eigensolution of $\bH^{\inf}$ is in general 
not consistent with the assumption of a steady state and the use of 
the popular NEGF+DFT formalism becomes questionable. 

We gratefully acknowledge financial support by the Deutsche
Forschungsgemeinschaft through the SFB 658, and the EU Network of Excellence
NANOQUANTA (NMP4-CT-2004-500198).


\begin{thebibliography}{23}
\expandafter\ifx\csname natexlab\endcsname\relax\def\natexlab#1{#1}\fi
\expandafter\ifx\csname bibnamefont\endcsname\relax
  \def\bibnamefont#1{#1}\fi
\expandafter\ifx\csname bibfnamefont\endcsname\relax
  \def\bibfnamefont#1{#1}\fi
\expandafter\ifx\csname citenamefont\endcsname\relax
  \def\citenamefont#1{#1}\fi
\expandafter\ifx\csname url\endcsname\relax
  \def\url#1{\texttt{#1}}\fi
\expandafter\ifx\csname urlprefix\endcsname\relax\def\urlprefix{URL }\fi
\providecommand{\bibinfo}[2]{#2}
\providecommand{\eprint}[2][]{\url{#2}}

\bibitem[{\citenamefont{{M.A.~Reed} et~al.}(1997)\citenamefont{{M.A.~Reed},
  Zhou, {C.J.~Muller}, {T.P.~Burgin}, and
  {J.M.~Tour}}}]{ReedZhouMullerBurginTour:97}
\bibinfo{author}{\bibnamefont{{M.A.~Reed}}},
  \bibinfo{author}{\bibfnamefont{C.}~\bibnamefont{Zhou}},
  \bibinfo{author}{\bibnamefont{{C.J.~Muller}}},
  \bibinfo{author}{\bibnamefont{{T.P.~Burgin}}}, \bibnamefont{and}
  \bibinfo{author}{\bibnamefont{{J.M.~Tour}}}, \bibinfo{journal}{Science}
  \textbf{\bibinfo{volume}{278}}, \bibinfo{pages}{252} (\bibinfo{year}{1997}).

\bibitem[{\citenamefont{{A.I.~Yanson} et~al.}(1998)\citenamefont{{A.I.~Yanson},
  {G.~Rubio~Bollinger}, {H.E.~van~den~Brom}, N.Agra{\"{\i}}t, and
  {J.M.~van~Ruitenbeek}}}]{YansonBollingerBromAgraitRuitenbeek:98}
\bibinfo{author}{\bibnamefont{{A.I.~Yanson}}},
  \bibinfo{author}{\bibnamefont{{G.~Rubio~Bollinger}}},
  \bibinfo{author}{\bibnamefont{{H.E.~van~den~Brom}}},
  \bibinfo{author}{\bibnamefont{N.Agra{\"{\i}}t}}, \bibnamefont{and}
  \bibinfo{author}{\bibnamefont{{J.M.~van~Ruitenbeek}}},
  \bibinfo{journal}{Nature} \textbf{\bibinfo{volume}{395}},
  \bibinfo{pages}{783} (\bibinfo{year}{1998}).

\bibitem[{\citenamefont{Meir and {N.S.~Wingreen}}(1992)}]{MeirWingreen:92}
\bibinfo{author}{\bibfnamefont{Y.}~\bibnamefont{Meir}} \bibnamefont{and}
  \bibinfo{author}{\bibnamefont{{N.S.~Wingreen}}}, \bibinfo{journal}{Phys.~Rev.
  Lett.} \textbf{\bibinfo{volume}{68}}, \bibinfo{pages}{2512}
  (\bibinfo{year}{1992}).

\bibitem[{\citenamefont{Landauer}(1957)}]{Landauer:57}
\bibinfo{author}{\bibfnamefont{R.}~\bibnamefont{Landauer}},
  \bibinfo{journal}{IBM~J.~Res.~Develop.} \textbf{\bibinfo{volume}{1}},
  \bibinfo{pages}{233} (\bibinfo{year}{1957}).

\bibitem[{\citenamefont{B{\"u}ttiker}(1986)}]{Buettiker:86}
\bibinfo{author}{\bibfnamefont{M.}~\bibnamefont{B{\"u}ttiker}},
  \bibinfo{journal}{Phys.~Rev. Lett.} \textbf{\bibinfo{volume}{57}},
  \bibinfo{pages}{1761} (\bibinfo{year}{1986}).

\bibitem[{\citenamefont{{N.D.~Lang}}(1995)}]{Lang:95}
\bibinfo{author}{\bibnamefont{{N.D.~Lang}}}, \bibinfo{journal}{Phys.~Rev.~B}
  \textbf{\bibinfo{volume}{52}}, \bibinfo{pages}{5335} (\bibinfo{year}{1995}).

\bibitem[{\citenamefont{Hirose and Tsukada}(1995)}]{ht.1995}
\bibinfo{author}{\bibfnamefont{K.}~\bibnamefont{Hirose}} \bibnamefont{and}
  \bibinfo{author}{\bibfnamefont{M.}~\bibnamefont{Tsukada}},
  \bibinfo{journal}{Phys.~Rev.~B} \textbf{\bibinfo{volume}{51}},
  \bibinfo{pages}{5278} (\bibinfo{year}{1995}).

\bibitem[{\citenamefont{{J.M.~Seminario}
  et~al.}(1998)\citenamefont{{J.M.~Seminario}, {A.G.~Zacarias}, and
  {J.M.~Tour}}}]{SeminarioZacariasTour:98}
\bibinfo{author}{\bibnamefont{{J.M.~Seminario}}},
  \bibinfo{author}{\bibnamefont{{A.G.~Zacarias}}}, \bibnamefont{and}
  \bibinfo{author}{\bibnamefont{{J.M.~Tour}}},
  \bibinfo{journal}{J.~Am.~Chem.~Soc.} \textbf{\bibinfo{volume}{120}},
  \bibinfo{pages}{3970} (\bibinfo{year}{1998}).

\bibitem[{\citenamefont{Taylor et~al.}(2001)\citenamefont{Taylor, Guo, and
  Wang}}]{tgw.2001}
\bibinfo{author}{\bibfnamefont{J.}~\bibnamefont{Taylor}},
  \bibinfo{author}{\bibfnamefont{H.}~\bibnamefont{Guo}}, \bibnamefont{and}
  \bibinfo{author}{\bibfnamefont{J.}~\bibnamefont{Wang}},
  \bibinfo{journal}{Phys.~Rev.~B} \textbf{\bibinfo{volume}{63}},
  \bibinfo{pages}{245407} (\bibinfo{year}{2001}).

\bibitem[{\citenamefont{Palacios et~al.}(2001)\citenamefont{Palacios,
  P\'erez-Jim\'enez, Louis, and Verg\'es}}]{pplv.2001}
\bibinfo{author}{\bibfnamefont{J.~J.} \bibnamefont{Palacios}},
  \bibinfo{author}{\bibfnamefont{A.~J.} \bibnamefont{P\'erez-Jim\'enez}},
  \bibinfo{author}{\bibfnamefont{E.}~\bibnamefont{Louis}}, \bibnamefont{and}
  \bibinfo{author}{\bibfnamefont{J.}~\bibnamefont{Verg\'es}},
  \bibinfo{journal}{Phys.~Rev.~B} \textbf{\bibinfo{volume}{64}},
  \bibinfo{pages}{115411} (\bibinfo{year}{2001}).

\bibitem[{\citenamefont{Xue et~al.}(2002)\citenamefont{Xue, Datta, and
  {M.A.~Ratner}}}]{XueDattaRatner:02}
\bibinfo{author}{\bibfnamefont{Y.}~\bibnamefont{Xue}},
  \bibinfo{author}{\bibfnamefont{S.}~\bibnamefont{Datta}}, \bibnamefont{and}
  \bibinfo{author}{\bibnamefont{{M.A.~Ratner}}}, \bibinfo{journal}{Chem.~Phys.}
  \textbf{\bibinfo{volume}{281}}, \bibinfo{pages}{151} (\bibinfo{year}{2002}).

\bibitem[{\citenamefont{Brandbyge et~al.}(2002)\citenamefont{Brandbyge,
  {J.-L.~Mozos}, Ordej\'{o}n, Taylor, and
  Stokbro}}]{BrandbygeMozosOrdejonTaylorStokbro:02}
\bibinfo{author}{\bibfnamefont{M.}~\bibnamefont{Brandbyge}},
  \bibinfo{author}{\bibnamefont{{J.-L.~Mozos}}},
  \bibinfo{author}{\bibfnamefont{P.}~\bibnamefont{Ordej\'{o}n}},
  \bibinfo{author}{\bibfnamefont{J.}~\bibnamefont{Taylor}}, \bibnamefont{and}
  \bibinfo{author}{\bibfnamefont{K.}~\bibnamefont{Stokbro}},
  \bibinfo{journal}{Phys.~Rev.~B} \textbf{\bibinfo{volume}{65}},
  \bibinfo{pages}{165401} (\bibinfo{year}{2002}).

\bibitem[{\citenamefont{Evers et~al.}(2004)\citenamefont{Evers, Weigend, and
  Koentopp}}]{EversWeigendKoentopp:04}
\bibinfo{author}{\bibfnamefont{F.}~\bibnamefont{Evers}},
  \bibinfo{author}{\bibfnamefont{F.}~\bibnamefont{Weigend}}, \bibnamefont{and}
  \bibinfo{author}{\bibfnamefont{M.}~\bibnamefont{Koentopp}},
  \bibinfo{journal}{Phys.~Rev.~B} \textbf{\bibinfo{volume}{69}},
  \bibinfo{pages}{235411} (\bibinfo{year}{2004}).

\bibitem[{\citenamefont{Faleev et~al.}(2005)\citenamefont{Faleev, Leonard,
  Stewart, and van Schilf\-gaarde}}]{flss.2005}
\bibinfo{author}{\bibfnamefont{S.~V.} \bibnamefont{Faleev}},
  \bibinfo{author}{\bibfnamefont{F.}~\bibnamefont{Leonard}},
  \bibinfo{author}{\bibfnamefont{D.~A.} \bibnamefont{Stewart}},
  \bibnamefont{and} \bibinfo{author}{\bibfnamefont{M.}~\bibnamefont{van
  Schilf\-gaarde}}, \bibinfo{journal}{Phys.~Rev.~B}
  \textbf{\bibinfo{volume}{71}}, \bibinfo{pages}{195422}
  (\bibinfo{year}{2005}).

\bibitem[{\citenamefont{Koentopp et~al.}(2008)\citenamefont{Koentopp, Chang,
  Burke, and Car}}]{KoentoppChangBurkeCar:08}
\bibinfo{author}{\bibfnamefont{M.}~\bibnamefont{Koentopp}},
  \bibinfo{author}{\bibfnamefont{C.}~\bibnamefont{Chang}},
  \bibinfo{author}{\bibfnamefont{K.}~\bibnamefont{Burke}}, \bibnamefont{and}
  \bibinfo{author}{\bibfnamefont{R.}~\bibnamefont{Car}},
  \bibinfo{journal}{J.~Phys.~Condens.~Matter} \textbf{\bibinfo{volume}{20}},
  \bibinfo{pages}{083203} (\bibinfo{year}{2008}).

\bibitem[{\citenamefont{Stefanucci and
  {C.-O.~Almbladh}}(2004{\natexlab{a}})}]{StefanucciAlmbladh:04}
\bibinfo{author}{\bibfnamefont{G.}~\bibnamefont{Stefanucci}} \bibnamefont{and}
  \bibinfo{author}{\bibnamefont{{C.-O.~Almbladh}}},
  \bibinfo{journal}{Europhys.~Lett.} \textbf{\bibinfo{volume}{67}},
  \bibinfo{pages}{14} (\bibinfo{year}{2004}{\natexlab{a}}).

\bibitem[{\citenamefont{Stefanucci and
  {C.-O.~Almbladh}}(2004{\natexlab{b}})}]{StefanucciAlmbladh:04-2}
\bibinfo{author}{\bibfnamefont{G.}~\bibnamefont{Stefanucci}} \bibnamefont{and}
  \bibinfo{author}{\bibnamefont{{C.-O.~Almbladh}}},
  \bibinfo{journal}{Phys.~Rev.~B} \textbf{\bibinfo{volume}{69}},
  \bibinfo{pages}{195318} (\bibinfo{year}{2004}{\natexlab{b}}).

\bibitem[{\citenamefont{Dhar and Sen}(2006)}]{DharSen:06}
\bibinfo{author}{\bibfnamefont{A.}~\bibnamefont{Dhar}} \bibnamefont{and}
  \bibinfo{author}{\bibfnamefont{D.}~\bibnamefont{Sen}},
  \bibinfo{journal}{Phys.~Rev.~B} \textbf{\bibinfo{volume}{73}},
  \bibinfo{pages}{085119} (\bibinfo{year}{2006}).

\bibitem[{\citenamefont{Stefanucci}(2007)}]{Stefanucci:07}
\bibinfo{author}{\bibfnamefont{G.}~\bibnamefont{Stefanucci}},
  \bibinfo{journal}{Phys.~Rev.~B} \textbf{\bibinfo{volume}{75}},
  \bibinfo{pages}{195115} (\bibinfo{year}{2007}).

\bibitem[{\citenamefont{Kurth et~al.}(2005)\citenamefont{Kurth, Stefanucci,
  {C.-O.~Almbladh}, Rubio, and
  {E.K.U.~Gross}}}]{KurthStefanucciAlmbladhRubioGross:05}
\bibinfo{author}{\bibfnamefont{S.}~\bibnamefont{Kurth}},
  \bibinfo{author}{\bibfnamefont{G.}~\bibnamefont{Stefanucci}},
  \bibinfo{author}{\bibnamefont{{C.-O.~Almbladh}}},
  \bibinfo{author}{\bibfnamefont{A.}~\bibnamefont{Rubio}}, \bibnamefont{and}
  \bibinfo{author}{\bibnamefont{{E.K.U.~Gross}}},
  \bibinfo{journal}{Phys.~Rev.~B} \textbf{\bibinfo{volume}{72}},
  \bibinfo{pages}{035308} (\bibinfo{year}{2005}).

\bibitem[{\citenamefont{{S.-H.~Ke} et~al.}(2004)\citenamefont{{S.-H.~Ke},
  Baranger, and Yang}}]{KeBarangerYang:04}
\bibinfo{author}{\bibnamefont{{S.-H.~Ke}}},
  \bibinfo{author}{\bibfnamefont{H.}~\bibnamefont{Baranger}}, \bibnamefont{and}
  \bibinfo{author}{\bibfnamefont{W.}~\bibnamefont{Yang}},
  \bibinfo{journal}{Phys.~Rev.~B} \textbf{\bibinfo{volume}{70}},
  \bibinfo{pages}{085410} (\bibinfo{year}{2004}).

\bibitem[{\citenamefont{Li et~al.}(2007)\citenamefont{Li, Zhang, Hou, Qian,
  Shen, Zhao, and Xue}}]{LiZhangHouQianShenZhaoXue:07}
\bibinfo{author}{\bibfnamefont{R.}~\bibnamefont{Li}},
  \bibinfo{author}{\bibfnamefont{J.}~\bibnamefont{Zhang}},
  \bibinfo{author}{\bibfnamefont{S.}~\bibnamefont{Hou}},
  \bibinfo{author}{\bibfnamefont{Z.}~\bibnamefont{Qian}},
  \bibinfo{author}{\bibfnamefont{Z.}~\bibnamefont{Shen}},
  \bibinfo{author}{\bibfnamefont{X.}~\bibnamefont{Zhao}}, \bibnamefont{and}
  \bibinfo{author}{\bibfnamefont{Z.}~\bibnamefont{Xue}},
  \bibinfo{journal}{Chem.~Phys.} \textbf{\bibinfo{volume}{336}},
  \bibinfo{pages}{127} (\bibinfo{year}{2007}).

\bibitem[{\citenamefont{Stefanucci et~al.}(2008)\citenamefont{Stefanucci,
  Kurth, Rubio, and {E.K.U.~Gross}}}]{StefanucciKurthRubioGross:08}
\bibinfo{author}{\bibfnamefont{G.}~\bibnamefont{Stefanucci}},
  \bibinfo{author}{\bibfnamefont{S.}~\bibnamefont{Kurth}},
  \bibinfo{author}{\bibfnamefont{A.}~\bibnamefont{Rubio}}, \bibnamefont{and}
  \bibinfo{author}{\bibnamefont{{E.K.U.~Gross}}},
  \bibinfo{journal}{Phys.~Rev.~B} \textbf{\bibinfo{volume}{77}},
  \bibinfo{pages}{075339} (\bibinfo{year}{2008}).

\end{thebibliography}
\end{document}